\documentclass{cmspaper}
\usepackage{graphicx}
\usepackage{here}
\usepackage{epsfig}

\begin{document}

\def \mbul {\noindent $\bullet$} 
\def \bd   {  $\diamond$}  
\newcommand{\br}[1]{\overline{#1}}
\def \et {\mbox{$E_T$}} 
\def \pt {\mbox{$p_T$}}
\def\etmin{\mbox{$E_T^{min}$}}


\begin{titlepage}

  \conferencereport{2006/016}
   \date{14 March 2006}

\title{Event generators for top quark production and decays}

  \begin{Authlist}
    S.R.~Slabospitsky 
       \Instfoot{ihep}{Institute for High Energy Physics, Protvino, Russia}
  \end{Authlist}

\collaboration{CMS collaboration}

  \begin{abstract}
 This contribution provides brief descriptions the 
current status  of the basic Monte Carlo event generators for
 top-quark  production and decays.
  \end{abstract} 

\conference{Presented at {\it International Workshop on Top Quark Physics},
  Coimbra, Portugal, January 12-15, 2006}
  
\end{titlepage}

\setcounter{page}{2}

\section*{INTRODUCTION } 

The Standard Model (SM) top quark couplings are uniquely fixed by the
principle of gauge invariance, the structure of the quark generations,
and a requirement of including the lowest dimension interaction
Lagrangian~\cite{Beneke:2000hk}.  Within the SM the top quark is considered as a 
point-like particle.
It should be stressed, that within the SM all top-quark production properties 
and decays are evaluated with high accuracy without any  phenomenological 
parameters. The total cross section production as well as the
differential distributions are calculated with ${\cal O}(10\%)$ 
accuracy~\cite{Beneke:2000hk}.
The top quark decays through one  decay channel, $t \to b W^+$ (other decay 
channels have very small branching ratios,  less then ${\cal O}(10^{-3}))$.
 Due to a very small life-time  
($\tau_t \sim 10^{-24}$~sec, $\tau_t \ll 1/\Lambda_{\rm QCD}$) the top-quark 
decays long before it can hadronize.  Therefore, it is unlikely to  expect
the formation of top-hadrons, $T(t \bar t)$- or $M(t \bar q)$-mesons and
$\Lambda(t q q)$-baryons.
The search for anomalous (i.e. non-SM) interactions is one of
the main motivations for studying top-quark physics.
In addition, very often the events with the top quarks are backgrounds to new
 physics that we hope to discover.

The physics of the top quark at the LHC  has been studied in 
great detail~\cite{Beneke:2000hk}, including in many cases a realistic simulation 
of the detectors  (see talks at this Workshop). The goal of
 this presentation is to give a short 
review of Monte Carlo (MC) generators that provide a simulation of the processes 
with top-quark production and decays (the detailed considerations could be found
elsewhere~\cite{Beneke:2000hk, Dobbs:2004qw}).

\section { GENERAL SCHEME OF GENERATOR  }

A general scheme of complete event simulation implies  the evaluation of the hard 
process (the cross section value, the incoming and outgoing particle's momenta and
 colours), then the evolution  the event through a parton showering and 
the hadronization of the coloured products of shower, 
followed by the decay of the unstable particles.
As a result the event is described by the momenta of the final hadrons,  leptons and 
photons and positions of their decay vertexes. Typically such information includes
 also the characteristics (momenta, colours, {\tt KF}-codes, mother's 
and daughter's relations) 
of  all intermediate partons (quarks, gluons, gauge bosons, unstable physical 
particles, etc)  that allows  to trace-back the history of particles production 
inside the event. Any such generator using an acceptance-rejection method 
(Von Neumann) returns weighted event. The most popular complete event generators,
like {\small HERWIG}~\cite{Corcella:2000bw}, {\small PYTHIA}~\cite{pythia},
{\small ISAJET}~\cite{isajet}, and {\small SHERPA}~\cite{sherpa} 
include all such steps of simulation.

\begin{figure}[!h]
\begin{center}
\includegraphics[width=0.5\textwidth,clip]{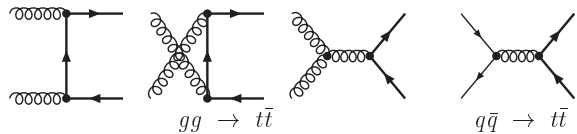} 
\caption{ $t \bar t$ production (QCD)  }
\label{fig:dia-ttbar}
\end{center}
\end{figure}

\begin{figure}[!h]
\begin{center}
\includegraphics[width=0.5\textwidth,clip]{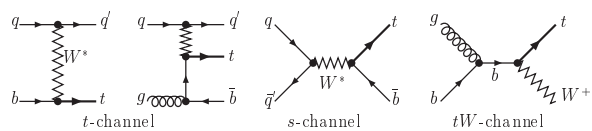} 
\caption{ single top production (electro-weak)  }
\label{fig:dia-sng}
\end{center}
\end{figure}

The processes with top-quarks and their realization in MC generators are described in 
the following.
The diagrams describing top-pair and single top production (at leading order -- LO)
are  presented on Figs~\ref{fig:dia-ttbar}, \ref{fig:dia-sng}.

In almost all cases one needs to know the information not only about the final 
top-quark, but also about its decay products. 
The natural and correct way to include decays is to calculate an exact amplitude 
with the initial partons and final state particles. However, the full amplitude 
typically consists  also of many diagrams without intermediate top-quarks. 
For examples, the full amplitude of the process $g g \to t \bar t$ 
with the both final $W$-bosons decaying into electrons has 
many other diagrams without $t$-quarks. 
In particular, there is at least one diagram with two $Z$-bosons: 
$g g \to b \bar b Z Z$ with one $Z$ decaying
into $e^+ e^-$ and second $Z \to \nu_e \bar \nu_e$. Typically,  all such
 non-resonant diagrams (i.e. without top-quark) will give a very small 
contribution to ``resonance'' region, where 
when the invariant masses of the final particles are very close to $t$-quark and 
$W$-boson masses
(e.g. $M(b e^{\pm} \nu_e) \approx m_t$ and $M(e^{\pm} \nu_e) \approx M_W$).
Therefore, for this case it is useful to introduce the  Narrow Width Approximation 
(NWA). It means
that one take into account only diagrams with intermediate top quarks and consider 
all $t$-quarks as ``on-shell'' particles:
\begin{eqnarray}
 \frac{1}{(p_t^2 - m_t^2)^2 + m^2_t \Gamma^2_t}
 = \frac{\pi}{ m_t \Gamma_t} \delta(p_t^2 - m_t^2) 
\label{eq:top1}
\end{eqnarray}

As a result one can factorize the production and decays of the top quark.
After that, typically there are  three important issues: 
i)~how to include top decays ?; ii)~how to reproduce Breit-Wigner resonance shape ?;
iii)~how to include top spin (polarization) ?

\section { TOP DECAYS  } 

\subsection { Decays  } 
Within the SM in 99.9\% of the cases the top-quark decays into the $b W$ final state:
 $t \to b W, \;\; W \to f \bar f'$. The matrix element describing this decay
is well known~\cite{Beneke:2000hk}:
\begin{eqnarray}
 |M(t \to b \, f \bar f')|^2 \propto 
   \frac{(p_b p_{f}) (p_{t} p_{\bar f'}) }
 {(p^2_{W} - M_W^2)^2 + \Gamma_W^2 M_W^2} 
\label{eq:top2}
\end{eqnarray}
\noindent 
In fact, this is the only one decay channel that is included in almost all generators.
Other interesting SM channels (e.g $t \to b W^* Z^* \, \to \, b f \bar f'
 \ell^+ \ell^-$) are still not available at generator level.
At the same time few other decay channels (due to non-SM physics) 
are included in generators. In particular, they are:

$\diamond$   SUSY: $t \to b H^+$ (almost all packages)

$\diamond$   FCNC: $t \to q g, \; q Z, \; q \gamma$ 
({\small TopReX}~\cite{Slabospitsky:2002ag})

$\diamond$ 
$t \to b W (\to f \bar f')$ with anomalous interactions 
($V+A$ and tensor couplings,  {\small ONETOP}~\cite{onetop})

\subsection { The Breit-Wigner resonance shape }

The  Narrow Width Approximation (NWA) assumes that all top quarks
 have the same default mass ($m_t$). There are two approaches, 
that allow  to reproduce the  Breit-Wigner resonance shape
due to the finite decay width of the top-quark.

With the ``smearing-mass'' method the matrix elements are 
 calculated for a default $m_{t}$ mass~\cite{pythia} . Then for each event 
the Breit-Wigner distribution
\begin{eqnarray}
 f(\tilde{m_t}) \propto    \frac{1}{(\tilde{m_t}^2 - m_t^2)^2 + \Gamma_t^2 m_t^2}
\label{eq:top4}
\end{eqnarray}
 is used to generate separate masses ($\tilde{m}_t$) for top-quarks.
The momenta and energies of all final-state particles are 
re-evaluated. For example, for $t \bar t$ pair production
with new $\tilde{m}_1(t)$ and  $\tilde{m}_2(\bar t)$ masses one has:
\begin{eqnarray}
t \bar t \;\; : \;\;
 E_1^*(t) = E_2^*(\bar t) = \sqrt{\hat s}/2
\; \Longrightarrow \;
 \tilde{E}_{1/2} = \frac{\sqrt{\hat s} \pm \tilde{m}^2_1(t) \mp
 \tilde{m}^2_2(\bar t)}{2 \sqrt{\hat s}}
\label{eq:mass}
\end{eqnarray}
This method certainly is not unique, but normally provides a sensible 
behavior. It can be used for event with any number of top quarks 
in final state ($t$, $t \bar t$, ...).

In another proposed method (modified NWA~\cite{onetop}) a 
 new $\tilde m_t$ for $t$-quark  in the event is generated by using of Breit-Wigner 
distribution.  Then, the squared matrix element ($|M|^2$) is calculated with 
this new $\tilde m_t$.
This method  can be used only for the process with one top quark in the 
final state (i.e. not for $t \bar t$ production).

\subsection { How to include top polarization }

Since the top quark decays before hadronization, its spin properties
are not spoiled. Therefore spin correlations in top production and
decays is an interesting issue in top-quark physics. 
The simplest way to include top polarization employs the helicity 
amplitude technique~\cite{Richardson:2001df}.
An equivalent (and sometimes more simple for practical usage)  
method~\cite{Jadach:1985ac} 
is realized in the {\small TAUOLA}~\cite{tauola}   package. 
In this method the matrix element squared $|M|^2$ can be represented in the
 ``factorized'' form.

\begin{figure}[!h]
\begin{center}
\includegraphics[width=0.3\textwidth,clip]{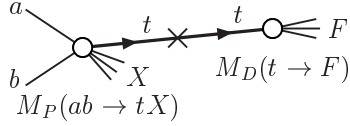}
\caption{ an equivalent method (TAUOLA)  }
\label{fig:tdec}
\end{center}
\end{figure}

For the process with one final top-quark (see Fig.~\ref{fig:tdec}) the full amplitude 
squared has the form:
\begin{eqnarray*}
&& |M(a b \to F + X)|^2 = \frac{\pi}{\Gamma_t m_t} \delta(p_t^2 - m_t^2) 
\times 
|M^{0}_{P}(a b \to t + X)|^2 \;\; { (1 + v_i h_i) } \;\;
|M^{0}_{D}(t \to F)|^2 \\ 
&& |M_{P}(a b \to t + X)|^2 = |M^{0}_{P}(a b \to t + X)|^2 (1 + (vs)) \\
&& |M_{D}(t \to F)|^2 = |M^{0}_{D}(t \to F)|^2 (1 + (hs))
\end{eqnarray*}
where the amplitudes  $|M^{0}|^2$ describe the production and  decay
of {\bf unpolarized} top-quark, $s$ is the $t$-quark spin ($(p_t s) = 0$),
and $v^{\mu}$, $h^{\mu}$ are the so-called ``polarization'' 
vectors~\cite{Jadach:1985ac}. 

For example, the spin-dependent matrix element squared describing top-quark decay
 $t \to b \ell^+ \nu$ can be re-presented in the form as follows:
\begin{eqnarray}
 |M|^2 \propto 
   \frac{(p_b p_{\nu}) (p_{t} p_{\ell}) }
 {(p^2_{W} - M_W^2)^2 + \Gamma_W^2 M_W^2} 
   \;\;     \left(1 - \frac{m_t (p_{\ell} s)}{ (p_{t} p_{\ell})} \right)
 \Rightarrow  v^{\mu} = -\frac{m_t p^{\mu}_{\ell}}{(p_t p_{\ell})}
 \; \Rightarrow \; -\vec n^{*}_{\ell}
\label{eq:dec1}
\end{eqnarray}
where  $\vec{n}^{*}_{\ell}$ is the direction
of $\ell^+$ momentum in $t$-quark rest frame.

\section {MATCHING ALGORITHMS}

\subsection {Matching Parton Showers and Matrix Elements}

Recently a substantial progress has been achieved in the simulation of processes with
 multi-jet final states~\cite{Mangano:2003ps,Hoche:2006ph}. 
Indeed, the description of multi-jets obtained
from the shower evolution is inaccurate, since hard radiation at large
angle is suppressed by the angular ordering prescription.
The available generators ({\small ALPGEN, CompHEP, Madevent}, etc) provide
a generation of top production with up to 6-jets~\cite{Mangano:2002ea, comphep,
Maltoni:2002qb, Stelzer:1994ta}. Due to the complexity
of the matrix elements(ME) evaluation for these many-body configurations, 
calculations are normally available only for LO cross sections.

Multi-parton events generated
using the exact leading order ME generator can be consistently evolved into
multi-jet final states by means of a shower MC. The main problem is how to eliminate
 the double counting,  where jets can arise from both the higher-order calculation 
and  from the   hard emission during the shower evolution.

There are several approaches to this problem, aiming at different levels of
 accuracy. The first one (``matrix-element correction'' technique~\cite{mec})
 corrects the approximate ME for the emission of the
 hardest gluon in a given process by using the exact LO ME.
The second is known as CKKW~\cite{ckkw}; 
its goal is to implement multi-jet ME  corrections at the leading, or 
next-to-leading 
logarithmic level. This method assumes a separation of the multi-jet phase-space into
the  domains covered by the ME calculation and the domains covered by the shower 
evolution.
Then by means of Sudakov re-weighting the ME's  weights to reproduce the 
probability of an 
exclusive  N-jet final state from the inclusive parton-level N-jet rate. 
This allows to add     parton-level event samples of different jet 
multiplicity.

\begin{figure}[!h]
\begin{center}
\includegraphics[width=0.5\textwidth,clip]{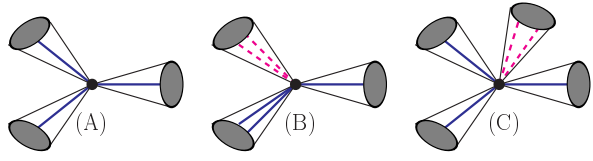}
\caption{Few examples of matching. The solid line corresponds to parton from a
 hard process, while the dashed one presents a parton emitted by the shower. 
(A) $N_{jet} = N_{\rm part} = 3$,
 the event is matched and kept. (B) $N_{jet} = N_{\rm part} = 3$, but 
$N_{\rm matched} = 2$, the event should be rejected.
(C) $N_{\rm matched} = N_{\rm part} = 3$, but
 $N_{jet} = 4$, the event is matched and kept for inclusive sample, but should
be rejected for exclusive samples. }
\label{fig:match}
\end{center}
\end{figure}

An additional prescription is proposed in~\cite{Mangano:2002ea} 
(``MLM prescription'').  In this approach the partons
from the ME calculation are matched to the jets reconstructed after the
perturbative shower.  Parton-level events are defined by a minimum
\et\ threshold \etmin\ for the partons, and a minimum separation among
them, $\Delta R_{jj}>R_{min}$.  
However, no Sudakov re-weighting is applied. Rather, events are showered, without
any hard-emission veto during the shower. After evolution, a jet cone
algorithm with cone size $R_{min}$ and minimum transverse energy
\etmin\ is applied to the final state. Starting from the hardest
parton, the jet which is closest to it in $(\eta,\phi)$ is selected.
If the distance between the parton and the jet centroid is smaller
than $R_{min}$, the parton and the jet match. The matched jet is
removed from the list of jets, and matching for subsequent partons is
performed. The event is fully matched if each parton has a matched
jet.  Events which do not match are rejected. 
For events
which satisfy matching, it is furthermore required that no extra jet,
in addition to those matching the partons, be present.  Events with
extra jets are rejected, a suppression replacing the Sudakov
re-weighting used in the CKKW approach. Events obtained by applying
this procedure to the parton level with increasing multiplicity can
then be combined to obtain fully inclusive samples spanning a large
multiplicity range. Events with extra jets are not rejected in the
case of the sample with highest partonic multiplicity. Fig.~\ref{fig:match}
presents an illustration of above-described 
examples~\cite{Mangano:2002ea, Hoche:2006ph}.

\subsection {Matching with and additional $b$-quark } 

\begin{figure}[!h]
\begin{center}
\includegraphics[width=0.3\textwidth,clip]{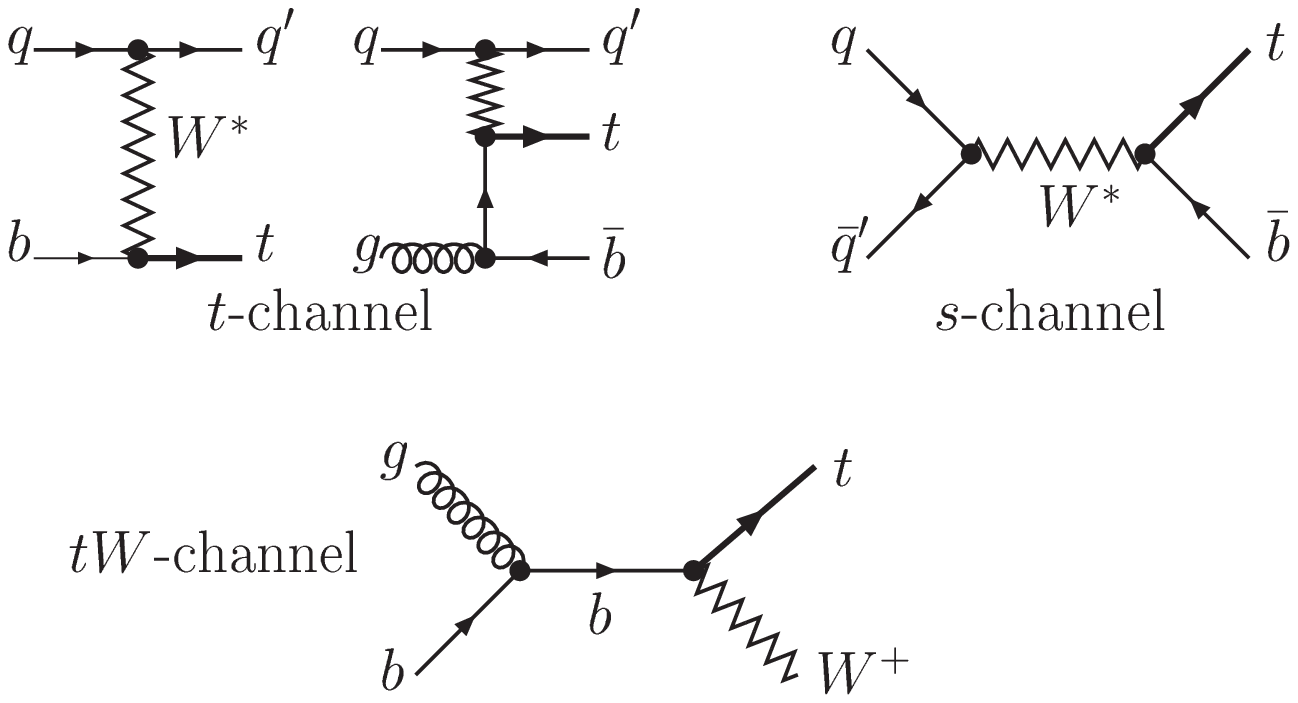} 
\caption{Two subprocess of the single top production.}
\label{fig:sng22}
\end{center}
\end{figure}

A very simple procedure is proposed for the process of $t$-channel single-top
production~\cite{singletop}, where one has two complementary 
subprocess (see Fig.~\ref{fig:sng22}):
\begin{eqnarray}
 ``2 \to 2'' q b \to q' t \quad {\rm and} \quad 
 ``2 \to 3'' q g \to q' t \bar b 
\label{sng2}
\end{eqnarray}

This matching procedure is based on the  transverse momentum 
($p_{\top}(\bar b)$) of additional $\bar b$-quark~\cite{singletop}.

\begin{figure}[!h]
\begin{center}
\includegraphics[width=10.truecm,height=4.truecm,clip=]{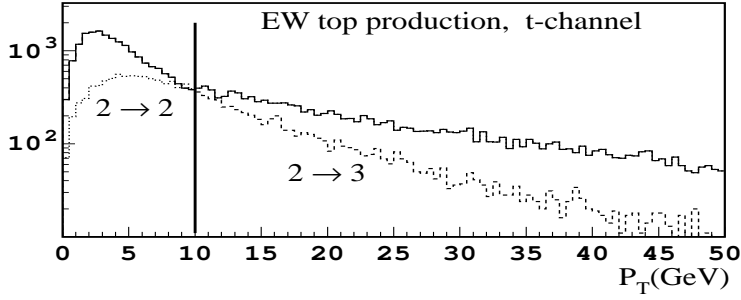} 
\caption{ $p_{\top}(\bar b)$  distributions
 of the additional $\bar b$-quark
 production in ``$2\to2$'' (the dotted curve) and ``$2\to3$'' (the dashed curve)
processes.
The solid histogram is the sum of these two contributions. 
 The vertical line corresponds
to parameter value  $p_0 = 10$~GeV. }  
\label{fig:pt-sng}
\end{center}
\end{figure}

The matching procedure starts from simulation of 
 the events with $t$-quark production due to  $2 \to 2$ process. Such an event
has an additional $\bar b$-quark generated by MC showering generator (i.e.
{\small PYTHIA} or {\small HERWIG}).  Any event from this sample is accepted 
if the transverse momentum of the additional $\bar b$-quark
 is smaller than some value  
$p_0$. The second sample consists of events from $2 \to 3$ 
process. The event from this sample is accepted if the additional $\bar b$-quark
from a hard process has $p_{\top}(\bar b) > p_0$. 
Thus, the resulting (total) sample of 
$t$-quark  production events is the sum of two contributions:
\begin{eqnarray}
N(p p \to t X)_{t-\hbox{channel}} &=& N^{(2\to 2)}(p p \to t q'; \,
p_{\top}(\bar b) < p_0) \label{dble}  \\
&+& N^{(2\to 3)}(p p \to t q' \bar b;  p_{\top}(\bar b) \geq p_0) 
\nonumber 
\end{eqnarray}
The corresponding $p_{\top}(\bar b)$-distributions are shown in 
Fig.~\ref{fig:pt-sng}.

This method of generation of $t$-channel single-top production
is realized  in {\small SingleTop}~\cite{singletop} and 
{\small TopReX}~\cite{Slabospitsky:2002ag} codes.

\section { NLO CORRECTIONS }

Recently a substantial progress has been achieved in the calculations of the 
single-top
production at NLO~\cite{ztop,onetop,campbell,mcnlo2}.
The radiation effects are included in the initial and final states, as well
as into decays. These generators (like {\small ZTOP}~\cite{ztop},
{\small ONETOP}~\cite{onetop} and the code based on 
{\small MCFM} generator~\cite{campbell}) provide a generation of total cross section
and differential distributions.

One really needs to include NLO corrections in ME generators.
Indeed,  only shower MC's provide a representation of the
final state complete enough to allow realistic detector
simulations.
  On the other hand, the inclusion of NLO effects in
fixed-order ME MC's leads to distributions which are not positive
definite. The naive
introduction of NLO matrix element could  lead to double counting, since 
 as shower MC generators already incorporate part of the NLO effects 
(a real emissions, as well
as virtual effects included in the Sudakov form factors).
This problem was successfully solved (see the article~\cite{Frixione:2002ik},
where it was shown  how this merging can be done very effectively). This method
is implemented in {\small MC@NLO} generator~\cite{mcnlo}
 describing heavy-quark pair, Higgs, DY and gauge boson pair production. 

\begin{figure}[h]
\begin{center}
\includegraphics[width=0.4\textwidth,clip=]{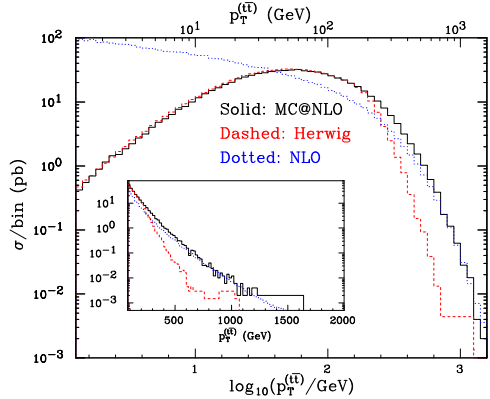}
\caption{Transverse momentum distribution of top quark pairs using
  three different approaches: the LO {\small HERWIG} MC, the parton-level NLO
  MC, and the merging of the two into {\small MC@NLO}. Figure
  from~\cite{Frixione:2003ei}. }
\label{fig:hwnlo}
\end{center}
\end{figure}

The inclusion of NLO corrections in the shower MC guarantees that
total cross-sections generated by the MC reproduce those of the NLO ME
calculation, thereby properly including the $K$ factors and
reducing the systematic uncertainties induced by renormalization and
factorization scale variations. At the same time, however, the
presence of the higher-order corrections generated by the shower will
improve the description of the NLO distributions, leading to
departures from the parton-level NLO result. This is shown, for
example, in Fig.~\ref{fig:hwnlo}, which shows the \pt\ spectrum of a
$t\bar{t}$ pair resulting from the pure NLO calculation, from the LO
shower, and from the {\small MC@NLO} improvement. At large \pt, a region
dominated by the NLO effects, {\small MC@NLO}  faithfully reproduces the hard,
large-angle emission distribution given by the NLO matrix elements. At
small \pt, a region dominated by multiple radiation and higher-order
effects, the {\small MC@NLO} departs significantly from the NLO result, while
properly incorporating the Sudakov re-summation effects only available
via the IS shower evolution. 

\begin{figure}[h]
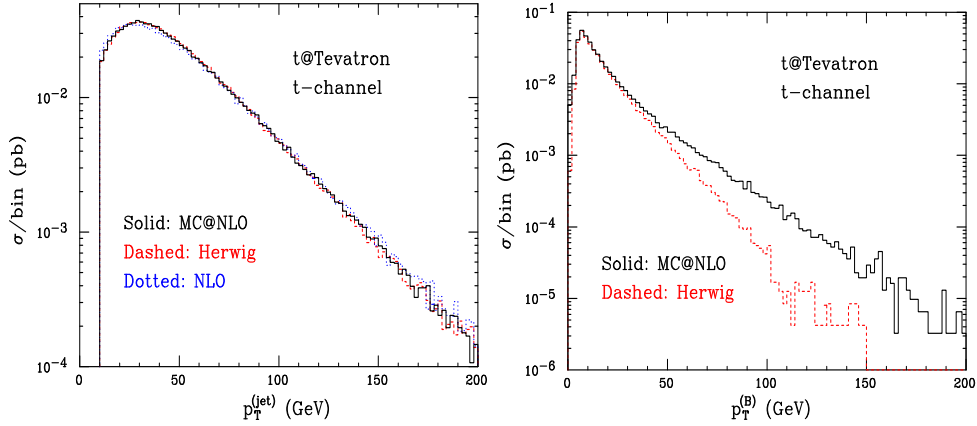

\begin{center}
\includegraphics[width=0.4\textwidth,height=0.35\textwidth,clip]{ptj1_tch_tev.eps} 
\includegraphics[width=0.4\textwidth,height=0.35\textwidth,clip]{ptb_tch_tev.eps} 
\caption{The transverse momentum distributions for the  highest $p_T$ jet
  and $b$-jet (not from top decay). }
\label{fig:nlo-pt}
\end{center}
\end{figure}

Recently, NLO corrections to the single-top production processes
 ($t$- and $s$-channels) are included
at {\small MC@NLO} event generator~\cite{mcnlo2}. The comparison {\small MC@NLO} and
HERWIG results for the single top production are given on Fig.~\ref{fig:nlo-pt}.
Note, that the both generators give very close results for the  highest $p_T$ jet 
distribution (the left histogram on  Fig.~\ref{fig:nlo-pt}), while {\small MC@NLO}
predicts much harder $p_T$-spectrum for $b$-jet (not from top decay).

\section {  EVENT GENERATORS }

This section presents a brief description of the generators. Only few basic comments
and the list of included top-quark production process are given
for each entry.

\subsection { Full event simulation packages }
These packages provide a full event simulation including the hard process generation, 
showering and hadronization with subsequent  decays of the unstable hadrons. 

\vspace{2mm} \mbul
 {\sf HERWIG}~\cite{Corcella:2000bw} \\
contains  a wide range of Standard Model, Higgs and 
supersymmetric processes. It uses the parton-shower approach for initial- 
and final-state QCD radiation, including colour coherence effects and 
azimuthal correlations both within and between the jets.
{\small HERWIG} is particularly sophisticated in its treatment of
  the subsequent decay of unstable resonances, including full spin
  correlations for most processes. \\
{\underline {\bf Processes included:}}  $t \bar t$, single top ($t$, $s$ channels),
 $t \bar t H$,   $Z t \bar t$, $g b \to t H^+$

\vspace{3mm} 
\mbul   {\sf PYTHIA }~\cite{pythia} \\  
is a general-purpose  generator for  hadronic events
in pp, e$^+$e$^-$ and $ep$ colliders. It contains a subprocess library
and generation machinery, initial- and final-state parton showers,
underlying event, hadronization and decays, and analysis tools. \\
{\underline {\bf Processes included:}} 
 $t \bar t$, single-top ($t$, $s$ channels),
  $t \bar t H$,
  $g b \to t H^+$, no spin correlations

\vspace{3mm }\mbul 
 {\sf ISAJET }~\cite{isajet} \\
 is a general-purpose  generator for  hadronic events.
{\small ISAJET} is based on perturbative QCD plus 
phenomenological models for parton and beam jet fragmentation. \\
\noindent{\underline {\bf Processes included:}}  $t \bar t$, no spin correlations

\vspace{2mm} \mbul 
 {\sf SHERPA }~\cite{sherpa} \\
 is a new multi purpose event generator with a powerful matrix element generator
{\small AMEGIC++}

\subsection { Tree level matrix element generators } 

Such packages generate the hard processes kinematic 
quantities, such as masses and momenta, the spin, the colour connection, and 
the flavour of initial- and final-state partons. Then such information
is stored in the ``Les Houches'' format~\cite{Boos:2001cv} 
 and is passed to full event simulation generator (like {\small PYTHIA} or 
{\small HERWIG}).

\vspace{2mm} \mbul 
 {\sf ALPGEN}~\cite{Mangano:2002ea} \\
 is designed for the generation of the SM processes 
in hadronic collisions, with emphasis on final states with large jet
multiplicities.  It is based on the exact LO evaluation of partonic matrix
elements, as well as top and gauge boson decays with helicity correlations. 
The code generates events in both a weighted and unweighted
mode. 

\noindent{\underline {\bf Processes included:}}  $t \bar t$ + up to 6jets,
single top: $tq$, $tb$, $tW$, $tbW$ (no extra jets), 
$t \bar t t \bar t$ + up to 4jets, $t \bar t b \bar b$ + up to 4jets,
$H t \bar t $ + up to 4jets,
$W/Z t \bar t $ + up to 4jets 

\vspace{2mm}\mbul 
 {\sf CompHEP}~\cite{comphep} \\ 
 {\small CompHEP} computes squared
Feynman diagrams symbolically and then numerically calculates cross
sections and distributions.  The output event's information
in the ``Les Houches'' format~\cite{Boos:2001cv}.

\noindent{\underline {\bf Processes included:}}  $t \bar t$,
single top, (?  $t \bar t t \bar t$, $t \bar t b \bar b$, )
$W/Z t \bar t$, spin correlations are included

\vspace{2mm} \mbul 
 {\sf  MadEvent }~\cite{Maltoni:2002qb} \\
is a multi-purpose, tree-level 
event generator,  which is powered by the matrix element generator 
{\small MadGraph}~\cite{Stelzer:1994ta}.
{\small MadGraph} automatically generates the amplitudes for all the relevant
subprocesses and produces the mappings for the integration over the
phase space.  

\noindent{\underline {\bf Processes included:}}  $t \bar t$ + up to 3jets,
single top,  $t \bar t b \bar b$ + up to 1jet,
$H t \bar t$ up to 2jets

\vspace{2mm} \mbul 
 {\sf MC@NLO }~\cite{mcnlo} 
combines a Monte Carlo event generator with exact NLO calculations of rates for 
QCD processes at hadron colliders.

\noindent{\underline {\bf Processes included:}}  $t \bar t$, 
single top ($t$- and $s$-channel)

\vspace{2mm}\mbul 
 {\sf AcerMC }~\cite{acermc}
(see talk of B.Krersevan in this conference). \\
{\small AcerMC} is dedicated for generation of the Standard Model background processes 
in $pp$ collisions at the LHC

\noindent{\underline {\bf Processes included:}} $t \bar t$,
single top, $t \bar t t \bar t$, $t \bar t b \bar b$,
$W/Z t \bar t$, spin correlations are included

\vspace{2mm}\mbul 
  {\sf SingleTop}~\cite{singletop} generator is based on the 
{\small CompHEP}~\cite{comphep} package.

\noindent {\underline {\bf processes:}} 
$t$-channel single top production ($2 \to 2$ + $2 \to 3$,
 spin correlations are included

\vspace{2mm} \mbul 
 {\sf TopReX}~\cite{Slabospitsky:2002ag} \\
provides a simulation of several important processes in 
hadronic collision, not implemented in {\small PYTHIA}. 
Several top-quark decays channels are included: the SM channel
($t \to q W^+$, $ q= d, s, b$), $b$-quark and charged Higgs
($t \to b H^+$) and the channels with flavor changing neutral current 
(FCNC): $t \to u(c) \; V$, $ V = g, \, \gamma, \, Z$.
The implemented matrix elements take into account  spin polarizations  of the
top quark.

\noindent{\underline {\bf Processes included:}} 
$gg\,(q \bar q) \to   t \bar t$, single-top production ($t$-, $s$-, and $tW$-channel),
 $q \bar q' \to H^{\pm*} \to t \bar b$, 
 $q \bar q \to W^{*}/Z^{*} Q \bar Q$ with $W^{*}/Z^{*} \to f \bar f$ and 
$Q = c, b, t$, 
 $g u(c) \to t \to b W$ (due to FCNC).

\vspace{2mm} \mbul 
 {\sf MCFM}~\cite{mcfm} 
included the matrix elements at next-to-leading order and incorporate full spin 
correlations.

\noindent{\underline {\bf Processes included:}}  $t \bar t$,
single top ($t$- and $s$-channel), $H t \bar t$, $W/Z t \bar t$

\vspace{2mm} \mbul 
 {\sf ZTOP}~\cite{ztop} code includes the full NLO-corrections to single top 
production ($t$- and $s$-channel). 

\vspace{2mm} \mbul 
 {\sf ONETOP}~\cite{onetop} code include 
full NLO-corrections to single top production
($t$- and $s$-channel) and top-quark decay.

\section* { CONCLUSIONS }

Recently  substantial progress has been achieved in the implementation of new ideas 
concerning top events simulation and in developing 
event generators for top quark production and decays.

In particular, several  new processes describing  top-quark production
and decays are included in the generators providing complete event simulation.
Many  event generators provide generation of top-quark processes with spin
 correlations.
Tree level generators make possible the generation of top  production processes
 with additional multi-jets in the final states. 
The codes with full NLO corrections to 
single-top production processes are available.
A few generators include also  processes with  top-quark
  production and decays due to interactions beyond SM.

\section* {Acknowledgments}
I would like to  thank the organizers of ``Top~2006'' 
for the invitation and extremely pleasant atmosphere and
fruitful discussion during the conference. 


\begin{thebibliography}{99}

\bibitem{Beneke:2000hk}
M.~Beneke {\it et al.},  \emph{ Top quark physics}, 
 [{\tt arXiv:hep-ph/0003033}],
in \emph{Standard model physics (and more) at the LHC} G.~Altarelli and
M.L.~Mangano eds.,
{\it  Geneva, Switzerland: CERN (2000) 529 p}.
 
\bibitem{Dobbs:2004qw}
M.~Dobbs  {\it et al.} \emph{Les Houches Guidebook to Monte Carlo Generators for 
Hadron Collider Physics}, {\tt hep-ph/0403045}

\bibitem{Corcella:2000bw}
G.~Corcella {\it et al.},
\emph{HERWIG 6: An event generator for hadron emission reactions with 
 interfering gluons (including supersymmetric processes)}, 
\emph{JHEP} {\bf 0101} (2001) 010 [{\tt arXiv:hep-ph/0011363}]. \\ 
{\underline{Webpage:}} {\tt http://hepwww.rl.ac.uk/theory/seymour/herwig}


\bibitem{pythia} 
T.~Sj\"ostrand {\it et al.}, \emph{High-energy-physics event generation with 
PYTHIA 6.1}, \emph{Comput.\ Phys.\ Commun.} {\bf 135} (2001) 238
 [{\tt arXiv:hep-ph/0010017}]. \\
{\underline{Webpage:}} {\tt http://www.thep.lu.se/$\sim$torbjorn/Pythia.html}

\bibitem{isajet}
  F.E.~Paige, S.D.~Protopopescu, H.~Baer and X.~Tata,
\emph{ISAJET~7.69: A Monte Carlo event generator for $p p$, $\bar p p$, and 
$e^+ e^-$ reactions}, {\tt arXiv:hep-ph/0312045}. \\
{\underline{Webpage:}} {\tt  http://www.phy.bnl.gov/$\sim$isajet/}

\bibitem{sherpa} 
  T.~Gleisberg, S.~Hoeche, F.~Krauss, A.~Schalicke, S.~Schumann and J.~Winter,
\emph{Event generator for the LHC}, {\tt arXiv:hep-ph/0508315}. \\
{\underline{Webpage:}} {\tt  http://www.physik.tu-dresden.de/$\sim$krauss/hep/ } 

\bibitem{Richardson:2001df}
  P.~Richardson,  \emph{Spin correlations in Monte Carlo simulations},
 \emph{JHEP} {\bf 0111} (2001) 029   [{\tt arXiv:hep-ph/0110108}].

\bibitem{Jadach:1985ac}
  S.~Jadach and Z.~Was,
\emph{QED ${\cal O}(\alpha^3)$ Radiative Corrections To The Reaction 
$e^= e^- \to \tau^+ \tau^-$ Including Spin And Mass Effects. (Erratum)},
\emph{Acta Phys.\ Polon.} {\bf B15} (1984) 1151 
  [Erratum-ibid.\ {\bf B16} (1985) 483].

\bibitem{tauola}
S.~Jadach, J.H.~Kuhn and Z.~Was,
 \emph{Tauola: A Library Of Monte Carlo Programs To Simulate Decays Of
Polarized Tau Leptons},  \emph{Comput.\ Phys.\ Commun.}  {\bf 64} (1990) 275. \\
{\underline{Webpage:}} {\tt http://wasm.home.cern.ch/wasm/goodies.html}

\bibitem{Mangano:2003ps}
  M.L.~Mangano,  \emph{QCD tools for the LHC},  \emph{eConf} {\bf C030614} (2003) 015
  [{\tt arXiv:hep-ph/0312117}].

 \bibitem{Mangano:2002ea}
 M.L.~Mangano, M.~Moretti, F.~Piccinini, R.~Pittau, and A.D.~Polosa,
 \emph{ALPGEN, a generator for hard multiparton processes in hadronic collisions},
 \emph{JHEP} {\bf 0307} (2003) 001  [{\tt arXiv:hep-ph/0206293}]. \\
{\underline{Webpage:}} {\tt http://mlm.home.cern.ch/mlm/alpgen/}


\bibitem{Hoche:2006ph}
 S.~Hoche, F.~Krauss, N.~Lavesson, L.~Lonnblad, M.~Mangano, A.~Schalicke, and 
S.~Schumann,
 \emph{Matching parton showers and matrix elements},   {\tt arXiv:hep-ph/0602031}.


\bibitem{comphep} 
E.~Boos, V.~Bunichev, M.~Dubinin, L.~Dudko, V.~Edneral, V.~Ilyin, A.~Kryukov,
V.~Savrin, A.~Semenov, A.~Sherstnev (the CompHEP collaboration) \\
{\underline{Webpage:}} {\tt http://theory.sinp.msu.ru/comphep}

\bibitem{Maltoni:2002qb}
F.~Maltoni and T.~Stelzer,
 \emph{MadEvent: Automatic event generation with MadGraph},
 \emph{JHEP} {\bf 0302} (2003) 027 [{\tt arXiv:hep-ph/0208156}]. \\
{\underline{Webpage:}} {\tt http://madgraph.physics.uiuc.edu}

\bibitem{Stelzer:1994ta}
T.~Stelzer and W.~F.~Long,
 \emph{Automatic generation of tree level helicity amplitudes}, 
 \emph{Comput.\ Phys.\ Commun.}  {\bf 81} (1994) 357 [{\tt arXiv:hep-ph/9401258}].

\bibitem{mec}
M.H.~Seymour,
 \emph{Matrix element corrections to parton shower algorithms},
 \emph{Comput.\ Phys.\ Commun.}  {\bf 90} (1995) 95 [{\tt hep-ph/9410414}].


\bibitem{ckkw}
  S.~Catani, F.~Krauss, R.~Kuhn and B.R.~Webber,
 \emph{QCD matrix elements + parton showers},  \emph{JHEP} {\bf 0111} (2001) 063
  [{\tt arXiv:hep-ph/0109231}]; \\
  L.~Lonnblad,
 \emph{Correcting the colour-dipole cascade model with fixed order matrix elements},
  \emph{JHEP} {\bf 0205} (2002) 046  [{\tt arXiv:hep-ph/0112284}].

\bibitem{Frixione:2002ik}
S.~Frixione and B.~R.~Webber,
 \emph{Matching NLO QCD computations and parton shower simulations},
   hep-ph/0204244; \\
S.~Frixione, P.~Nason and B.R.~Webber,
 \emph{Matching NLO QCD and parton showers in heavy flavour production},
  \emph{JHEP} {\bf 0308} (2003) 007 [{\tt hep-ph/0305252}].



\bibitem{singletop} 
 E.~Boos, L.~Dudko, V.~Savrin, CMS Note 2000/065 (2000).


\bibitem{Slabospitsky:2002ag}
S.R.~Slabospitsky and L.~Sonnenschein,
\emph{TopReX generator (version 3.25): Short manual},
\emph{ Comput.\ Phys.\ Commun.} {\bf 148} (2002) 87 (2002) 
[{\tt arXiv:hep-ph/0201292}]. \\
{\underline{Webpage:}} {\tt http://cmsdoc.cern.ch/$\sim$spitsky/toprex/toprex.html}


\bibitem{acermc}
B.P.~Kersevan and E.~Richter-Was,
\emph{The Monte Carlo event generator AcerMC version 2.0 with interfaces to
 PYTHIA~6.2 and HERWIG~6.5}, 
 {\tt arXiv:hep-ph/0405247}; \\
 B.P.~Kersevan and E.~Richter-Was,
  \emph{The Monte Carlo event generator AcerMC version 1.0 with interfaces to
  PYTHIA~6.2 and HERWIG~6.3},
 \emph{Comput.\ Phys.\ Commun.} {\bf 149} (2003) 142 [{\tt arXiv:hep-ph/0201302}]. \\
{\underline{Webpage:}} {\tt  http://borut.home.cern.ch/borut/}

\bibitem{Frixione:2003ei}
S.~Frixione, P.~Nason and B.~R.~Webber,
\emph{Matching NLO QCD and parton showers in heavy flavour production},
 \emph{JHEP} {\bf 0308} (2003) 007 [{\tt hep-ph/0305252}].

\bibitem{mcnlo} 
  S.~Frixione and B.R.~Webber,
\emph{The MC@NLO 3.2 event generator}, 
  {\tt arXiv:hep-ph/0601192}. \\
{\underline{Webpage:}} {\tt http://www.hep.phy.cam.ac.uk/theory/webber/MCatNLO}

\bibitem{ztop}  B.W.~Harris, E.~Laenen, L.~Phaf, Z.~Sullivan and S.~Weinzierl,
\emph{The fully differential single top quark cross section in  next-to-leading
order QCD}, 
\emph{Phys.\ Rev.}  {\bf D66} (2002) 054024 [{\tt arXiv:hep-ph/0207055}]; \\
 Z.~Sullivan, 
 \emph{Understanding single-top-quark production and jets at hadron colliders}, 
 \emph{Phys.\ Rev.} {\bf D70} (2004) 114012   [{\tt arXiv:hep-ph/0408049}].

\bibitem{onetop} 
 Q.H.~Cao, R.~Schwienhorst, J.A.~Benitez, R.~Brock, and C.-P.~Yuan,
  \emph{Next-to-leading order corrections to single top quark production and  decay
 at the Tevatron. II: t-channel process}, 
   \emph{Phys.\ Rev.} {\bf D72} (2005) 094027 [{\tt arXiv:hep-ph/0504230}]; \\
  Q.H.~Cao, R.~Schwienhorst and C.-P.~Yuan,
 \emph{Next-to-leading order corrections to single top quark production and  decay
  at Tevatron. I: s-channel process}, 
  \emph{Phys.\ Rev.} {\bf D71} (2005) 054023 [{\tt arXiv:hep-ph/0409040}]; \\
 Q.H.~Cao and C.-P.~Yuan,
\emph{Single top quark production and decay at next-to-leading order in hadron 
 collision}, 
  \emph{Phys.\ Rev.} {\bf D71} (2005) 054022  [{\tt arXiv:hep-ph/0408180}].

\bibitem{campbell}
  J.~Campbell, R.K.~Ellis and F.~Tramontano,
\emph{Single top production and decay at next-to-leading order}, 
\emph{Phys.\ Rev.} {\bf D70} (2004) 094012 [{\tt arXiv:hep-ph/0408158}].

\bibitem{mcnlo2} 
  S.~Frixione, E.~Laenen, P.~Motylinski and B.R.~Webber,
  \emph{Single-top production in MC@NLO}, {\tt arXiv:hep-ph/0512250}.

\bibitem{mcfm} J.~Campbell and K.~Ellis, 
\emph{{\small MCFM} -- Monte Carlo for FeMtobarn processes},  \\
{\underline{Webpage:}} {\tt http://mcfm.fnal.gov/}

\bibitem{Boos:2001cv}
E.~Boos {\it et al.}, \emph{Generic user process interface for event
generators}, {\tt arXiv:hep-ph/0109068}.

\end{thebibliography}
\end{document}